\documentclass[prd,twocolumn,showpacs,amsmath,amssymb,superscriptaddress,preprintnumbers,nofootinbib,showkeys]{revtex4}
\usepackage{amssymb}
 \usepackage{amsmath}
\usepackage{graphicx}
 \usepackage{euscript}

\newcommand{\nablab}{{\mathop {\rule{0pt}{0pt}{\nabla}}\limits^{\bot}}\rule{0pt}{0pt}}

\begin{document}

\title{Dynamic aether as a trigger for  spontaneous spinorization in early Universe}

\author{Alexander B. Balakin}
\email{Alexander.Balakin@kpfu.ru} \affiliation{Department of
General Relativity and Gravitation, Institute of Physics, Kazan
Federal University, Kremlevskaya str. 16a, Kazan 420008, Russia}

\author{Anna O. Efremova}
\email{anna.efremova131@yandex.ru} \affiliation{Department of
General Relativity and Gravitation, Institute of Physics, Kazan
Federal University, Kremlevskaya str. 16a, Kazan 420008, Russia}

\date{\today}

\begin{abstract}
In the framework of the Einstein-Dirac-aether theory we consider a phenomenological model of the spontaneous growth of the fermion number, which is triggered by the dynamic aether.
The trigger version of spinorization of the early Universe is associated with two mechanisms: the first one is the aetheric regulation of behavior of the spinor field; the second mechanism  can be related to a self-similarity of internal interactions in the spinor field.  The dynamic aether is designed to switch on and switch off the self-similar mechanism of the spinor field evolution; from the mathematical point of view, the key of such a guidance is made of the scalar of expansion of the aether flow, proportional to the Hubble function in the isotropic cosmological model. Two phenomenological parameters of the presented model are shown to be considered as factors predetermining the total number of fermions born in the early Universe.

\end{abstract}
\maketitle

\section{Introduction}

In 1996 Damour and Esposito-Far\`ese have introduced the term "spontaneous scalarization" \cite{SS0} in order to describe gravitational analogs of the phase transition of the second order in
ferromagnetic materials. The phenomenological idea about spontaneous scalarization has been used in different astrophysical and cosmological contexts (see, e.g.,  \cite{SS1}-\cite{SS12}). This fruitful idea  has been extended and applied to the models with other fields, and now we can find works
devoted to the problems of spontaneous vectorization (see, e.g., \cite{SV1,SV2,SV3}), spontaneous tensorization \cite{ST}, spontaneous spinorization \cite{SSpin0,SSpin1}, as well as, to the problems of spontaneous polarization of the
color aether \cite{CA1,CA2,CA3} and of spontaneous growth of the gauge fields \cite{SGG}.

The formalism of the spontaneous growth of the mentioned physical fields is mainly connected with the mechanism of  tachyonic instability.
We consider the phenomenon of the spontaneous spinorization, but propose another mechanism based on the model of self-similarity of the internal interactions in the fermion systems. Below we will discuss in detail this mechanism, but now we would like to focus on a new detail of our approach. We consider the spinor field in the framework of Einstein-Dirac-aether theory, and assume that the unit timelike vector field $U^j$ associated with the velocity four-vector of the dynamic aether is the key element of this theory. The theory of the dynamic aether (see, e.g., \cite{J1,J2,J3,J4,J5,J6,J7,J8} for
basic definitions and references) belongs to the category of vector-tensor modifications of gravity \cite{Odin1,Odin2}. The presence of the unit vector field $U^j$ in this theory
realizes the idea of a preferred frame of reference (see, e.g., \cite{CW,N1,N2,N3}), and indicates the possibility of violation of the Lorentz invariance \cite{LV1,LV2,LV3}.

Of course, it is hard to dispute the argument that the birth of particles is the field of the quantum theory. But the quantum version of the aetheric vector field is not yet established, and there are no
ideas what particles could be the carriers of the corresponding interactions. That is why, we restrict ourselves by the phenomenological theory. In fact, we consider some macroscopic consequences
of the interaction between the spinor and aether vector fields in order to formulate a hypothesis: when and how the spinorization provoked by the aether could happen in the early Universe.

What new detail does the involvement of the dynamic aether bring to the scheme of self-interaction of the spinor system? We assume that the aether regulates the dynamics of the fermion system.
What is the instrument of the aetheric influence on the spinor field? We assume that the key instrument of such guidance is the expansion scalar $\Theta = \nabla_k U^k$. On the one hand, this true scalar
is an intrinsic element of the aether flow. On the other hand, in the isotropic cosmological models of the Friedmann type $\Theta=3H$, where $H(t)= \frac{\dot{a}}{a}$ is the Hubble function. In other words, the scalar $\frac{3}{\Theta} = \frac{1}{H}$ defines the typical time scale, which characterizes the rate of the Universe evolution. Such a measure plays in the field theory the role analogous to the role of temperature, when one describes the Universe evolution on the thermodynamic level. This analogy is consistent with the fact that both quantities: the effective temperature and the expansion scalar are decreasing in the expanding Universe. But when we consider analogies between the expansion scalar $\Theta$ in the field theory and the temperature $T$ in the Universe thermodynamics, we can try to establish the analogy between the Curie temperature $T_{\rm C}$ in the theory of phase transitions of the second kind, and some critical
value $\Theta_{*}$. Such an approach allows us to suppose the following: A new internal interaction in the fermion system is switching on, if $\Theta<\Theta_{*}$ just like the phase transition
in ferroelectrics takes place, if $T$ becomes less than the Curie temperature $T<T_{\rm C}$. To conclude, we suppose that the aetheric guidance is manifested in the fact that the aether is switching
on (and switching off) the specific internal interaction in the fermion system in the manner of how the decreasing temperature switches on the reconstruction of ferromagnetic materials below the
Curie temperature. Our purpose is to show that such a mechanism could explain the spontaneous growth of the spinor particle number in early Universe.

The paper is organized as follows. In Section II, we reconstruct the total Lagrangian of the model, and derive the extended master equations for the unit vector, spinor and gravitational fields.
In Section III we consider the application to the isotropic cosmological model and derive the evolutionary equations for basic spinor invariants. In Section IV we present the model function describing the self-similar interaction in the fermion system and analyze the solutions of the corresponding extended master equations. Section V includes discussion and conclusions.

\section{The formalism}

\subsection{Lagrangian of the Einstein-Dirac-aether theory}

The canonic Lagrangian the Einstein-aether theory
$$
L_{({\rm EA})}= - \frac{1}{2\kappa}\left[R{+}2\Lambda + \lambda \left(g_{mn}U^m
U^n {-}1 \right) {+} \right.
$$
\begin{equation}
\left. + K^{abmn} \nabla_a U_m  \nabla_b U_n  \right]
\label{01}
\end{equation}
contains three principal parts \cite{J1}. In the first one $R$ is the Ricci
scalar, $\Lambda$ is the cosmological constant, and $\kappa {=} 8\pi G$ includes the Newtonian coupling constant $G$ ($c{=}1$). The second and third parts of the Lagrangian (\ref{01}) contain the four-vector $U^i$, associated with the aether velocity. The term $\lambda \left(g_{mn}U^m U^n {-}1 \right)$ designed to guarantee that the $U^i$ is normalized to one; respectively, $\lambda$ is the Lagrange multiplier.
The so-called kinetic term $K^{abmn} \ \nabla_a U_m \ \nabla_b U_n $ is quadratic in the covariant derivative
$\nabla_a U_m $ of the vector field $U^i$, with the tensor $K^{abmn}$ to be constructed
using the metric tensor $g^{ij}$ and the aether velocity four-vector $U^k$ only,
\begin{equation}
K^{abmn} {=} C_1 g^{ab} g^{mn} {+} C_2 g^{am}g^{bn}
{+} C_3 g^{an}g^{bm} {+} C_4 U^{a} U^{b}g^{mn}.
\label{2}
\end{equation}
The parameters $C_1$, $C_2$, $C_3$ and $C_4$ are the Jacobson coupling constants.
The massive spinor field is described by the following term of the Lagrangian:
\begin{equation}
L_{(\rm D)}=\frac{i}{2}[\bar{\psi}\gamma^{k}D_{k}\psi-D_{k}\bar{\psi}\gamma^{k}\psi]-m\bar{\psi}\psi  \,.
\label{Lspinor}
\end{equation}
Here $\psi$ defines the Dirac spinor field, $\bar{\psi}$ is the Dirac conjugated field;  $m$ is the mass prescribed to the spinor particle; $\gamma^k$ are the Dirac matrices, and the covariant (extended) derivatives of the spinors
\begin{equation}
D_{k}\psi = \partial_{k}\psi-\Gamma_{k}\psi \,, \quad
D_{k}\bar{\psi}=\partial_{k}\bar{\psi}+\bar{\psi}\Gamma_{k} \,,
\label{D}
\end{equation}
are constructed using the Fock-Ivanenko connection matrices $\Gamma_{k}$ \cite{FI}.

If we intend to construct the action functional of a multi-component system, we have to obey the following rules.
First, the  SU(N) symmetric Yang-Mills fields and thus the U(1) symmetric Maxwell field add the contributions of the form $-\frac14 {\bf F}_{mn}{\bf F}^{mn}$ (see, e.g., \cite{G}). Second, the scalar field introduces the term $\frac12 \left[ \nabla_m \phi \nabla^m \phi - m^2 \phi^2 \right]$. Third, the spinor field adds the term (\ref{Lspinor}), so that the Ricci scalar $R$, the invariant $\frac14 {\bf F}_{mn}{\bf F}^{mn}$, the  Klein-Gordon mass term $m^2\phi^2$ and the spinor mass term $m\bar{\psi}\psi$ enter the action functional with the same sign minus.
Taking into account this detail we use in our work the following total action functional
$$
- S_{({\rm EDA})} = \int d^4x \sqrt{-g}\left\{\frac{1}{2\kappa}\left[R{+}2\Lambda {+} \lambda \left(g_{mn}U^m
U^n {-}1 \right){+} \right. \right.
$$
$$
\left. \left.
+ K^{abmn} \nabla_a U_m  \nabla_b U_n  \right] + L_{(\rm matter)} + \beta(\Theta, S, P^2) - \right.
$$
\begin{equation}
\left. - \left[\frac{i}{2}[\bar{\psi}\gamma^{k}D_{k}\psi-D_{k}\bar{\psi}\gamma^{k}\psi]-m\bar{\psi}\psi \right] \right\} \,.
\label{0}
\end{equation}
The term $L_{(\rm matter)}$ describes the matter of non-spinor (non-fermionic) origin, e.g., the pseudo-Goldstone bosons attributed to the axionic dark matter. We include the minus sign into the left-hand side of this formula keeping in mind that the variation procedure $\delta S_{({\rm EDA})} = 0 = - \delta S_{({\rm EDA})}$ gives the same master equations. In addition,  we include into (\ref{0}) the cross term $\beta(\Theta, S, P^2)$; the arguments of this function are described below.

\subsection{Basic assumptions and auxiliary definitions}

\subsubsection{Fock-Ivanenko connection, tetrad four-vectors, spinor scalar $S$ and pseudoscalar $P$}

The Fock-Ivanenko matrices
\begin{equation}
\Gamma_{k}=\frac{1}{4}g_{mn}X^{(a)}_{s}\gamma^{s}\gamma^{n} \nabla_{k}X^{m}_{(a)}
\label{5D}
\end{equation}
contain four tetrad four-vectors $X^{m}_{(a)}$, which satisfy the relationships
\begin{equation}
g_{mn}X^{m}_{(a)}X^{n}_{(b)}=\eta_{(a)(b)} \,, \quad \eta^{(a)(b)}X^{m}_{(a)}X^{n}_{(b)}=g^{mn} \,,
\end{equation}
with the Minkowski metric $\eta_{(a)(b)}$.
The convolutions $\gamma^{k} = X^k_{(a)}\gamma^{(a)}$ links the Dirac matrices $\gamma^{k}$ depending on coordinates with the constant Dirac matrices $\gamma^{(a)}$.
As usual, the Dirac matrices satisfy the fundamental anti-commutation relations
$$
\gamma^{(a)} \gamma^{(b)} {+} \gamma^{(b)}\gamma^{(a)} =2 E \eta^{(a)(b)}  \,,
$$
\begin{equation}
\gamma^m \gamma^n   {+}  \gamma^n \gamma^m =2 E g^{mn} \,,
\label{D7}
\end{equation}
where $E$ is the unit matrix. Also, we keep in mind the formula
\begin{equation}
\epsilon_{mnpq} X^m_{(a)} X^n_{(b)} X^p_{(c)}X^q_{(d)} = \epsilon_{(a)(b)(c)(d)}\,,
\label{0D99}
\end{equation}
where $\epsilon_{mnpq}$ is the Levi-Civita tensor expressed via the absolutely antisymmetric symbol $E_{mnpq}$ as follows:
\begin{equation}
\epsilon_{mnpq} = \sqrt{-g} E_{mnpq} \,, \quad E_{0123} = -1 \,.
\label{Levi}
\end{equation}
In the Minkowski spacetime $\sqrt{-g}=1$, thus, $\epsilon_{(a)(b)(c)(d)} \equiv E_{(a)(b)(c)(d)}$ with $E_{(0)(1)(2)(3)}=-1$.
Using (\ref{0D99}) we can introduce in the covariant way the link between the Dirac matrices $\gamma^5$ and $\gamma^{(5)}$.
Indeed, according to the basic definition
\begin{equation}
\gamma^5 = -\frac{1}{4!} \epsilon_{mnpq} \gamma^m \gamma^n \gamma^p \gamma^q \,,
\label{gamma95}
\end{equation}
we obtain
$$
\gamma^5 = -\frac{1}{4!} \epsilon_{mnpq} X^m_{(a)} X^n_{(b)} X^p_{(c)}X^q_{(d)} \gamma^{(a)} \gamma^{(b)} \gamma^{(c)} \gamma^{(d)} =
$$
\begin{equation}
{=} {-}\frac{1}{4!} \epsilon_{(a)(b)(c)(d)} \gamma^{(a)} \gamma^{(b)} \gamma^{(c)} \gamma^{(d)}  {=}
\gamma^{(0)} \gamma^{(1)}\gamma^{(2)}\gamma^{(3)} \equiv \gamma^{(5)}.
\label{gamma85}
\end{equation}
In other words, the matrix $\gamma^{5}$ defined by (\ref{gamma95}) does not depend on metric and in addition to the unit matrix $E$ is a constant matrix.
This fact allows us to introduce the scalar  $S \equiv \bar{\psi}\psi= \bar{\psi}E \psi$ and the pseudoscalar $P \equiv i \bar{\psi}\gamma^5 \psi = i \bar{\psi}\gamma^{(5)} \psi$.
The scalar $S$ is usually associated with the density of the spinor particle number. As for the definition of $P$, the multiplier $i$ in front  provides the matrix $i\gamma^5$ to be free of the  imaginary unit.

\subsubsection{Decomposition of the covariant derivative of the aether velocity four-vector}

The tensor $\nabla_i U_k$ has the following standard decomposition into irreducible parts
\begin{equation}
\nabla_i U_k = U_i {\cal D}U_k + \sigma_{ik} + \omega_{ik} +
\frac{1}{3} \Delta_{ik} \Theta \,. \label{act3}
\end{equation}
Here ${\cal D}U^{i}$ is the acceleration four-vector, $\sigma_{ik}$ is
the shear tensor, $\omega_{ik}$ is the vorticity tensor, $\Theta$ is
the expansion scalar, $\Delta$ is the projector and ${\cal D}$ is the convective derivative:
$$
{\cal D}U_k \equiv  U^m \nabla_m U_k \,, \quad \sigma_{ik}
\equiv \frac{1}{2}\left(\nablab_i U_k {+}
\nablab_k U_i \right) {-} \frac{1}{3}\Delta_{ik} \Theta  \,,
$$
$$
\omega_{ik} \equiv \frac{1}{2} \left(\nablab_i U_k {-} \nablab_k U_i \right) \,, \quad \Theta \equiv \nabla_m U^m
\,,
$$
\begin{equation}
{\cal D} \equiv U^i \nabla_i \,, \quad \Delta^i_k = \delta^i_k - U^iU_k \,, \quad \nablab_i \equiv \Delta_i^k \nabla_k \,.
\label{act4}
\end{equation}
Using the presented decomposition one can say that there is one fundamental scalar $\Theta$ linear in the derivative, and three additional quadratic scalars associated with the aether flow:
\begin{equation}
a^2 \equiv {\cal D}U_k {\cal D}U^k \,, \quad \sigma^2 \equiv \sigma_{ik}\sigma^{ik}\,, \quad \omega^2 \equiv \omega_{ik}\omega^{ik} \,.
\label{b4}
\end{equation}
In the model studied below we use the new function $\beta(\Theta, S, P^2)$ of three arguments only, however, we hope to extend this modeling in the next works.

\subsection{Master equations}

The variation procedure with respect to the Lagrange multiplier $\lambda$, aether velocity four-vector $U^i$, spinor field $\psi$ and its Dirac conjugate quantity $\bar{\psi}$, and with respect to the metric $g^{pq}$ gives us  the coupled system of master equations of the model. We start with the derivation of the aether dynamic equations.

\vspace{3mm}
\subsubsection{Master equations for the aether velocity}

Variation of the total action functional (\ref{0}) with respect to the Lagrange multiplier $\lambda$ gives the condition
$g_{ik}U^i U^k {=} 1$. Variation with respect to $U^j$ yields
\begin{equation}
\nabla_{a}J^{aj}=I^j+\lambda U^j- \kappa \nabla^{j}\left(\frac{\partial \beta}{\partial \Theta}\right) \,,
\label{q1}
\end{equation}
where the terms ${\cal J}^{aj}$ and $I^j$ are defined as follows:
\begin{equation}
{\cal J}^{aj} = K^{abjn} (\nabla_b U_n) \,, \quad I^j =  C_4 ({\cal D}U_m)(\nabla^j U^m) \,,
\label{J7}
\end{equation}
and the Lagrange multiplier can be obtained as
\begin{equation}
\lambda = U_j \left[\nabla_{a}J^{aj}-I^j \right] + \kappa D\left(\frac{\partial \beta}{\partial \Theta}\right) \,.
\label{q2}
\end{equation}

\subsubsection{Master equations for the spinor field}

Variation with respect to $\bar{\psi}$ and $\psi$ gives, correspondingly
\begin{equation}
i\gamma^{n} D_{n}\psi =  M \psi \,, \quad i D_{n}\bar\psi\gamma^{n} = -\bar{\psi} M  \,,
\label{D1}
\end{equation}
\begin{equation}
M \equiv \left(m + \frac{\partial \beta}{\partial S}\right) E + \frac{\partial \beta}{\partial P} i \gamma^5 \,.
\label{M}
\end{equation}
Based on the matrix $M$ we can introduce the effective mass of the interacting spinor field
\begin{equation}
<M>  \equiv \frac{\bar{\psi}M \psi}{\bar{\psi} \psi} = \left(m + \frac{\partial \beta}{\partial S}\right)  + \left(\frac{P}{S}\right)\frac{\partial \beta}{\partial P} \,.
\label{D17}
\end{equation}

\subsubsection{Master equations for the gravity field}

Variation with respect to metric yields
\begin{equation}
R_{pq}{-}\frac12 g_{pq} R  {=} \Lambda g_{pq} {+} T^{(\rm U)}_{pq} {+} \kappa \left[T^{(\rm D)}_{pq} {+} T^{(\rm M)}_{pq} {+} T^{(\rm C)}_{pq}\right] ,
\label{D2}
\end{equation}
where the following terms reconstruct the total stress-energy tensor
$$
T^{(\rm U)}_{pq} =
\frac12 g_{pq} \ K^{abmn} \nabla_a U_m \nabla_b U_n{+} \lambda U_pU_q {+}
$$
$$
+C_1\left[(\nabla_mU_p)(\nabla^m U_q) {-}
(\nabla_p U_m )(\nabla_q U^m) \right] {+}C_4 {\cal D} U_p {\cal D} U_q +
$$
\begin{equation}
{+}\nabla^m \left[U_{(p}{\cal J}_{q)m} {-}
{\cal J}_{m(p}U_{q)} {-}
{\cal J}_{(pq)} U_m\right] \,,
\label{5Ein1}
\end{equation}
$$
T^{(\rm D)}_{pq} = - g_{pq} L_{D}  +
$$
\begin{equation}
 +  \frac{i}{4}\left[\bar\psi \gamma_{p} D_{q}\psi {+} \bar\psi \gamma_{q} D_{p}\psi {-} (D_{p}\bar\psi) \gamma_{q} \psi {-} (D_{p}\bar\psi) \gamma_{q} \psi \right] \,.
\label{D99}
\end{equation}
The contribution of the non-spinor matter $T^{(\rm M)}_{pq}$ is described by the standard formula
\begin{equation}
T^{(\rm M)}_{pq} = - \frac{2}{\sqrt{-g}} \frac{\delta}{\delta g^{pq}}\left[\sqrt{-g} L_{(\rm matter)} \right]  \,.
\label{D77}
\end{equation}
The cross-term $T^{(\rm C)}_{pq}$ associated with the function $\beta(\Theta, S, P^2)$ is of the form
\begin{equation}
T^{(\rm C)}_{pq} = g_{pq}\left[ \beta - \left(D + \Theta \right)\frac{\partial \beta}{\partial \Theta} \right]  \,.
\label{D777}
\end{equation}
Using the Dirac equations (\ref{D1}) one can obtain that
\begin{equation}
L_{D} = \left[ \bar\psi(M {-} mE) \psi \right]  = \left[S \frac{\partial \beta}{\partial S} {+} P \frac{\partial \beta}{\partial P}\right]\,.
\label{D994}
\end{equation}
We used the following auxiliary formulas for the variation of tetrad four-vectors with respect to metric:
\begin{equation}
\delta X^j_{(a)} = \frac14 \left[X_{p(a)}\delta^j_{q} + X_{q(a)}\delta^j_{p} \right]\delta g^{pq} \,,
\label{D764}
\end{equation}
(see, e.g., \cite{B1,B2} for details). Also, we used the rules
$$
\delta U^j = 0 \,, \quad  \delta \gamma^{(5)} = 0 = \delta \gamma^{5} \,,
$$
\begin{equation}
\delta \gamma^{(a)} = 0 \,, \quad \delta \gamma^{k} = \gamma^{(a)} \delta X^{k}_{(a)}  \,.
\label{D774}
\end{equation}

\section{Cosmological application}

\subsection{Geometrical aspects of the model}

For investigation of the spinorization phenomenon we consider the spatially isotropic homogeneous spacetime platform with the FLRW type metric
\begin{equation}
ds^2 = dt^2 - a^2(t)[{dx^1}^2 + {dx^2}^2 + {dx^3}^2]
\label{FLRW1}
\end{equation}
with the scale factor $a(t)$. For such a symmetry the aether velocity four-vector has to be of the form $U^j = \delta^j_0$, and the covariant derivative is simplified essentially:
\begin{equation}
\nabla_k U^m = H(t)\left(\delta^m_1 \delta_k^1 + \delta^m_2 \delta_k^2 + \delta^m_3 \delta_k^3 \right) \,,
\label{nablaU}
\end{equation}
where $H(t) \equiv \frac{\dot{a}}{a}$ is the Hubble function (here and below the dot symbolizes the derivative with respect to time).
The corresponding acceleration four-vector, shear tensor and vorticity tensor vanish, and the expansion scalar $\Theta {=} \nabla_k U^k$ is equal to $\Theta {=} 3H$.

The sum of the Jacobson coupling constants $C_1{+}C_3$ has been estimated in 2017 as the result of observation of the binary neutron star merger (the events GW170817 and GRB 170817A \cite{170817}).
It was established that the ratio of the  velocities of the gravitational and electromagnetic waves satisfies the inequalities $1-3 \times 10^{-15}< \frac{v_{\rm gw}}{c}<1+ 7 \times 10^{-16} $).
According to \cite{J2}  the square of the velocity of the tensorial aether mode is equal to  $S^2_{(2)}= \frac{1}{1{-}(C_1{+}C_3)}$, thus, the sum of the parameters $C_1{+}C_3$ can
be estimated as $-6 \times 10^{-15}<C_1{+}C_3< 1.4 \times 10^{-15}$. Clearly, we can consider that with very high precision $C_3 {=} -C_1$.
The parameter $C_4$ does not enter the key formulas since $DU^j=0$ for the FLRW model. As for the parameters $C_1$ and $C_2$, the results of the discussion about their constraints
\cite{CH1,CH2,constr1,constr2}) allows us to use the estimation $-\frac{2}{27} <C_2< \frac{2}{21}$ (see \cite{Dark}).
Taking into account these details we see that the tensor $J^{aj}$ is symmetric, and its nonzero components can be written as follows:
\begin{equation}
 J^a_j = C_2 \Theta   \delta^a_j  \,.
\label{JFRiedmann}
\end{equation}
The equations for the unit vector field (\ref{q1}) convert now into one equation
\begin{equation}
 C_2 D \Theta + \kappa D \left(\frac{\partial \beta}{\partial \Theta} \right)= \lambda   \,,
\label{reduceJac}
\end{equation}
which gives, in fact, the solution for the Lagrange multiplier $\lambda$.

Our supplementary assumption is that the non-spinor matter is a cold dust, and its stress-energy tensor is divergence-free, $\nabla^q T^{(\rm M)}_{pq}{=}0$, providing that $T^{(\rm M)}_{pq} {=} \rho  U_p U_q$ ($\rho$ is the corresponding energy density scalar). Then, the equations for the gravitational field can be reduced to the following one equation
\begin{equation}
\frac{1}{\kappa}\left[3H^2 \Gamma {-} \Lambda \right] = \left(\rho + m S \right) + \beta - \Theta \frac{\partial \beta}{\partial \Theta} \,.
\label{keyGravity}
\end{equation}
Here we introduced a new auxiliary parameter:
\begin{equation}
\Gamma = 1 + \frac32 C_2 \,.
\label{GammaC}
\end{equation}
Other Einstein's equations are the differential consequences of the evolutionary equations for the aether and spinor fields.
As a consequence of the separate conservation law for the non-fermionic matter we obtain the standard law of its evolution
\begin{equation}
\dot{\rho} + 3 H \rho = 0 \ \Rightarrow \ \rho(t) = \rho(t_0) \frac{a^3(t_0)}{a^3(t)} \,.
\label{Gamma99}
\end{equation}
For the metric (\ref{FLRW1}) the set of the tetrad four-vectors is
\begin{equation}
X^i_{(0)} = U^i = \delta^i_0 \,, \quad X^i_{(\alpha)} = \delta^i_{\alpha} \frac{1}{a(t)} \,, \quad (\alpha = 1,2,3) \,,
\label{tetradF}
\end{equation}
and the spinor connection coefficients (\ref{5D}) have the form
\begin{equation}
\Gamma_0 = 0 \,, \quad \Gamma_{\alpha} = \frac12 \dot{a} \gamma^{(\alpha)} \gamma^{(0)} \,.
\label{GFriedmann}
\end{equation}
We also use the direct consequence of (\ref{GFriedmann})
\begin{equation}
\gamma^{k} \Gamma_k = - \frac32 H \gamma^0 = -  \Gamma_k \gamma^{k} \,.
\label{conseq}
\end{equation}

\subsection{Reduced evolutionary equation for the spinor field}

We assume that the components of the spinor field are the functions of the cosmological time only; then the Dirac equations (\ref{D1}) yield
\begin{equation}
i \gamma^{0}\left(\partial_0 + \frac32 H \right) \psi = M \psi \,,
\label{DF1}
\end{equation}
\begin{equation}
i \left(\partial_0 + \frac32 H \right) \bar{\psi} \gamma^0 = - \bar{\psi} M \,.
\label{DF2}
\end{equation}
If one uses the replacement
\begin{equation}
\psi = a^{-\frac32} \Psi \,, \quad \bar{\psi} = a^{-\frac32} \bar{\Psi} \,,
\label{reopace}
\end{equation}
the Dirac equations take the form
\begin{equation}
i\gamma^{(0)} \dot{\Psi} = M \Psi \,, \quad i \dot{\bar{\Psi}} \gamma^{(0)} = - \bar{\Psi} M  \,.
\label{keyD}
\end{equation}

\subsection{Evolution of the spinor invariants}

Keeping in mind the equations (\ref{keyD}), we can find the rates of evolution of the invariants $S$ and $P$.
First, we see that
$$
\frac{d}{dt}\left(\bar{\psi} \psi \right) =
\frac{d}{dt}\left(\bar{\Psi} a^{-3}\Psi \right) =
$$
$$
= {-}3H \left(\bar{\psi} \psi \right) {+} a^{-3}\left[\left(\frac{d}{dt}\bar{\Psi}\right) {\gamma^{(0)}}^2 \Psi {+} \bar{\Psi}{\gamma^{(0)}}^2 \left(\frac{d}{dt} \Psi\right) \right]{=}
$$
\begin{equation}
= {-}3H \left(\bar{\psi} \psi \right) {+} i \bar{\psi} \left(M\gamma^0 {-} \gamma^0 M \right)\psi \,.
\label{dotS1}
\end{equation}
Using (\ref{M}) we can  present the evolutionary equation for the scalar $S$ as follows:
\begin{equation}
\dot{S} + 3H S = - 2 T \ \frac{\partial \beta}{\partial P} \,.
\label{dotS2}
\end{equation}
The evolutionary equation for the pseudoinvariant $P$ is
\begin{equation}
\dot{P} + 3H P = - \bar{\psi} \left(M\gamma^0 \gamma^5 {-} \gamma^5 \gamma^0 M \right)\psi \,,
\label{0dotP}
\end{equation}
\begin{equation}
\dot{P} + 3H P =  2 T \left( \frac{\partial \beta}{\partial S} + m \right) \,,
\label{dotP}
\end{equation}
where the following  auxiliary function $T$ is introduced:
\begin{equation}
T \equiv  \bar{\psi}\gamma^5 \gamma^0 \psi \,.
\label{T1}
\end{equation}
The auxiliary function $T(t)$ itself satisfies the equation
\begin{equation}
\dot{T} + 3H  T = -i \bar{\psi} \left(M \gamma^5 + \gamma^5 M \right)\psi \,,
\label{dotT}
\end{equation}
or equivalently
\begin{equation}
\dot{T} + 3H T  = -2 P \left(m + \frac{\partial \beta}{\partial S}\right) - 2 S \frac{\partial \beta}{\partial P}  \,.
\label{dotT2}
\end{equation}
In other words, the set of functions $S(t)$, $P(t)$ and $T(t)$ forms the closed evolutionary system.

One can explicitly check that the set of equations (\ref{dotS2}), (\ref{dotP}) and (\ref{dotT2}) for arbitrary $\beta$ admits the so-called first integral.
Indeed, the direct differentiation gives
\begin{equation}
\frac{d}{dt} \left(S^2-P^2-T^2\right) + 6H \left(S^2-P^2 -T^2\right) = 0 \,,
\label{dotT21}
\end{equation}
providing that
\begin{equation}
S^2-P^2-T^2 = \frac{{\rm const}}{a^6} \,.
\label{dotT22}
\end{equation}
For further progress it is convenient to introduce the variable
$x= \frac{a(t)}{a(t_0)}$, the dimensionless scale factor, where $t_0$ is some fixed moment of the cosmological time. Also, we introduce three auxiliary functions of this variable:
\begin{equation}
X = x^3 S \,, \quad Y = x^3 P \,, \quad Z = x^3 T \,,
\label{dotT27}
\end{equation}
so that the first integral (\ref{dotT22}) takes the form
\begin{equation}
X^2-Y^2-Z^2 = K
\label{dotT233}
\end{equation}
with arbitrary constant of integration $K$.
In these terms the evolutionary equations for $X(x)$ and $Y(x)$ take, respectively, the forms:
\begin{equation}
X^{\prime}(x) = - \frac{6}{x\Theta} Z \left[\frac{\partial \beta}{\partial P} \right] \,,
\label{X1}
\end{equation}
\begin{equation}
Y^{\prime}(x) =  \frac{6}{x\Theta} Z \left[m + \frac{\partial \beta}{\partial S} \right] \,,
\label{Y1}
\end{equation}
where $Z$ has to be extracted from (\ref{dotT233}), i.e.,
\begin{equation}
Z = \pm \sqrt{X^2-Y^2- K}
\label{Z1}
\end{equation}
with positive, negative or vanishing  parameter $K$.

1. When $K$ is positive, i.e., $K{=}\nu^2 >0$, the parametrization of the relationship (\ref{dotT233}) is
\begin{equation}
X = \nu \cosh{u(x)} \,, \quad  Y = \nu \sinh{u(x)} \cos{v(x)} \,,
\label{N1}
\end{equation}
$$
Z = \nu \sinh{u(x)} \sin{v(x)} \,,
$$
where $u(x)$ and $v(x)$ are real functions.

2. When $K$ is negative, i.e., $K{=} {-} \mu^2 <0$, we have the following parametrization   of  (\ref{dotT233}):
\begin{equation}
X = \mu \sinh{u(x)} \,, \quad  Y = \mu \cosh{u(x)}\cos{u(x)}  \,,
\label{N2}
\end{equation}
$$
Z = \mu \cosh{u(x)} \sin{v(x)} \,.
$$

3. When $K{=}0$, we deal, respectively, with the parametrization
\begin{equation}
X = u(x)  \,, \  Y =  u(x) \cos{v(x)} \,, \ Z = u(x) \sin{v(x)} \,.
\label{N3}
\end{equation}
In the first and second cases we deal with the hyperbolic laws of evolution of the function $X$ describing the number density of spinor particles.

\section{Modeling of the function $\beta(\Theta,S,P^2)$}

We would like to mention that new contributions to the Lagrangian, which have the form $F(S,P^2)$, were already considered in the nonlinear versions of the Einstein-Dirac models
(see, e.g., \cite{Saha1,Saha2,Saha3}). Also, the models describing the interaction of the spinor and scalar fields \cite{Saha5}, as well as, the spinor and pseudoscalar (axion) fields
\cite{BaEf} have been studied. We introduce the new element of the Lagrangian  $\beta(\Theta,S,P^2)$, which depends on the scalar of expansion of the aether flow $\Theta = \nabla_k U^k$.
Our ansatz is that the scalar $\beta$ as a function of the expansion scalar $\Theta$ is step-like:
\begin{equation}
\beta(\Theta,S,P^2) = \eta(\Theta_* {-}\Theta) \ \eta(\Theta {-} \Theta_{**}) B(S,P^2) \,,
\label{step1}
\end{equation}
where $\eta({\cal F})$ is the Heaviside function, which is equal to zero, if ${\cal F}<0$, and is equal to one, if ${\cal F}>0$.
This means that there exists some moment of the cosmological time $t_*$, when the interaction, described by the function $\beta$,
switches on. At this moment we fix the dimensionless scale factor $x(t_*) \equiv x_*$, and the corresponding values $\Theta_* \equiv \Theta(t_*)$ and $H_* \equiv H(t_*)$.
Similarly, we introduce the time moment $t_{**}$, when this interaction switches out. In other words, the aether flow guides the evolution of the spinor field.
From the mathematical point of view, one has to solve the set of evolutionary equations in three domains: first, when $t_0<t<t_*$; second, when $t_*<t<t_{**}$, third, when $t>t_{**}$. We assume that namely the second time interval relates to the spinorization phenomenon. At the borders $t{=}t_{*}$ and $t{=}t_{**}$ the functions $B(t)$, $S(t)$, $P(t)$ and $H(t)$ are assumed to be continuous.
We use in our assumptions the analogy with ferroelectric materials, which possess a pair of Curie temperatures $T_{\rm C1}$ and $T_{\rm C2}$.
Let us start the analysis for the first indicated interval.

 \subsection{Solutions for the interval $t_0 <t<t_*$ ($1<x<x_*$)}

 When $\beta =0$, we see from (\ref{X1}) that
 \begin{equation}
 X^{\prime}=0 \Rightarrow S(x)= \frac{S(1)}{x^3} \,,
 \label{step01}
 \end{equation}
 where $S(1) \equiv S(x(t_0))$ is the starting density of number of the spinor particles. Keeping in mind (\ref{Gamma99}) we find from the equation (\ref{keyGravity})                                                                                                                                                                                 the following Hubble function (in terms of $x$):
\begin{equation}
H(x) = \pm \sqrt{\frac{\Lambda}{3\Gamma}} \sqrt{1 + \frac{\kappa}{\Lambda x^3} \left[\rho(1)+ m S(1)\right]}  \,.
\label{kG2}
\end{equation}
For the description of the expanding Universe we have to choose the sign plus in (\ref{kG2}) and obtain that the function $H(x)$ is monotonic and is falling, i.e.,
$H^{\prime}(x) <0$. The prime symbolizes the derivative with respect to $x$. We see that $H_* < H(1)$ and thus $\Theta_* < \Theta(1)$.
Keeping in mind the relationship
\begin{equation}
t-t_0 = \int_1^{\frac{a(t)}{a(t_0)}} \frac{dx}{x H(x)} \,,
\label{kG3}
\end{equation}
we obtain the following result of integration:
$$
\frac{a(t)}{a(t_0)} = \left\{\cosh{\left[\sqrt{\frac{3\Lambda}{4\Gamma}}(t{-}t_0) \right]} {+} \right.
$$
\begin{equation}
\left. +\sqrt{1{+}\frac{\kappa }{\Lambda}\left[\rho(t_0) + m S(t_0) \right]} \ \sinh{\left[\sqrt{\frac{3\Lambda}{4\Gamma}}(t{-}t_0) \right]} \right\}^{\frac23} \,.
\label{kG5}
\end{equation}
Respectively, the effective mass coincides with the standard one, $<M>=m$.

\subsection{Solutions for the interval $t_* <t<t_{**}$ ($x_*<x<x_{**}$)}

\subsubsection{Hypothesis of self-similarity }

We assume that the function $B(S,P^2)$ has a self-similar form
\begin{equation}
 B(S,P^2) = S F(Q) \,, \quad Q = \frac{P^2}{S^2} \,,
\label{SS1}
\end{equation}
and obtain that the effective mass (\ref{D17}) takes the form
\begin{equation}
<M> = m + F(Q) \,.
\label{mass333}
\end{equation}
Combining the equations (\ref{X1}) and (\ref{Y1}) we obtain
\begin{equation}
 \frac{dY}{dX} = \frac{Y}{X} - \frac{X}{Y} \left[\frac{m + F(Q)}{2 F^{\prime}(Q)} \right] \,,
\label{SS13}
\end{equation}
or equivalently
\begin{equation}
 X \frac{dQ}{dX} = - \left[\frac{m + F(Q)}{2 F^{\prime}(Q)} \right] \,.
\label{SS17}
\end{equation}
The solution to this equation is
\begin{equation}
F(Q) = - m + \frac{{\rm const}}{X} \,,
\label{SS18}
\end{equation}
providing that
\begin{equation}
B(S,P^2) = SF(Q) = - m S(x) + \frac{{\rm const}}{x^3} \,.
\label{B77}
\end{equation}
We can find the constant of integration in (\ref{B77}) using the continuity requirements
\begin{equation}
B(x_*)=0 \,, \quad \lim_{x \to x_*-0} S(x) = \lim_{x \to x_*+0} S(x) \,.
\label{XY18}
\end{equation}
Clearly, we obtain
\begin{equation}
{\rm const} = m x^3_{*} S(x^*) = m S(1) \,.
\label{const1}
\end{equation}
This means that the right-hand side of the gravity field equation (\ref{keyGravity})
\begin{equation}
\rho {+} m S {+} SF(Q) {-} \Theta \delta(\Theta{-}\Theta_*) S F(Q) = \frac{\rho(1)}{x^3} {+}  \frac{m S(1)}{x^3}
\label{2keyGravity}
\end{equation}
keeps the same form as at $t_0 <t<t_*$. In other words, the solution for the scale factor coincides by the form with (\ref{kG5}),
but we have to replace $t_0$ with $t_*$, describing the starting point of the arguments of hyperbolic functions in the second time interval.

The function $Y^2(x)$ can be now presented as
\begin{equation}
Y^2 = X^2 F^{-1}\left[- m + \frac{m S(1)}{X}\right]
\label{2SS18}
\end{equation}
in terms of the inverse function $F^{-1}$. Thus, we need to solve the equation (\ref{X1}) and to extract the fermion density number function $S(x)= \frac{X(x)}{x^3}$. This task can be solved only numerically, that is why to have some analytical progress we consider below the linear function $F(Q) \to - m_* + h_0 Q$.

\subsubsection{Linear function $F(Q)$}

Let us  consider the function $B(S,P^2)$ in the form
\begin{equation}
B(S,P^2) = -m_* S + h_0 \frac{P^2}{S} \,,
\label{B1}
\end{equation}
where $m_*$ and $h_0$ are some phenomenological parameters.
We obtain now
\begin{equation}
Q(X) = \xi + \frac{m S(1)}{h_0 X} \,,
\label{XY5}
\end{equation}
where a new guiding parameter $\xi$ appears
\begin{equation}
\xi = \left(\frac{m_*{-}m}{h_0}\right) \,.
\label{XY3}
\end{equation}
Now the equation (\ref{X1}) can be rewritten as follows:
\begin{equation}
\frac{dX}{dx} = \mp \frac{4h_0 \sqrt{Q}}{x H} \sqrt{X^2(1-Q) - K} \,.
\label{XY6}
\end{equation}
Clearly, this equation admits separation of variables
\begin{equation}
\frac{\sqrt{X}dX}{\sqrt{\xi X + \frac{m S(1)}{h_0}} \sqrt{X \left[ \left(1-\xi \right) X - \frac{m S(1)}{h_0} \right] - K}} = \mp 4h_0 dt \,,
\label{XY9}
\end{equation}
but further analytic progress is possible for specific choice of the guiding parameters $\xi$ and $K$. Below we consider two such special cases.

\subsubsection{The submodel with $\xi=0$ and $K=0$}

When $K{=}0$, we work with the function $Z {=} \pm \sqrt{X^2{-}Y^2}$, and when $\xi =0$ we see that $m_*=m$.
For this choice of the guiding parameters we obtain the exactly integrable submodel with
\begin{equation}
X(t) = \frac{m S(1)}{h_0} \left[1+ 4h_0^2 \left(t-t_{+} \right)^2 \right] \,,
\label{XY11}
\end{equation}
\begin{equation}
Y(t) = \frac{m S(1)}{h_0} \sqrt{1+ 4h_0^2 \left(t - t_{+} \right)^2 } \,.
\label{XY12}
\end{equation}
Here we used the new auxiliary quantity
\begin{equation}
t_{+} = t_* + \sqrt{\frac{h_0-m}{4m h^2_0}} \,. \label{XY14}
\end{equation}
Also we used the boundary condition $X(t_*)= x^3_* S(x_*) = S(1)$, we have chosen the upper sign in (\ref{XY9}) so that $t_{+} > t_*$, and have assumed that $h_0 \geq m$.

The effective mass depends on time as  follows:
\begin{equation}
<M> = h_0 Q(t) = \frac{h_0}{\left[1{+} 4h_0^2 \left(t{-}t_{+} \right)^2 \right]} \,.
\label{M33}
\end{equation}
The function $<M>(t)$ starts with the value $m$ at $t=t_*$, then  reaches the maximal value $<M>_{(\rm max)} {=} h_0 \geq m$ at the moment $t= t_{\rm max}=t_{+}$, and then it decreases monotonically. The time moment $t_{\rm max}=t_{+}$ is the function of the parameter $h_0$ (see (\ref{XY14})); when $h_0=m$, $t_{\rm max} = t_{*}$; when $ m<h_0<2m$ this function grows and reaches the maximum at $h_0= 2m$; when $h_0 > 2m$ the parameter $t_{\rm max}$ decreases and tends to $t_*$. In other words, when the guiding parameter $h_0$ monotonically grows, the maximum of the function $<M>(t)$ drifts to the late time moments, then stops and starts to drift to the initial point $t_*$. Fig.1 illustrates the behavior of the reduced function $\frac{<M>}{m}$ for the cases, when $h_0 \geq 2$.

\begin{figure}
\includegraphics[width=85mm,height=70mm]{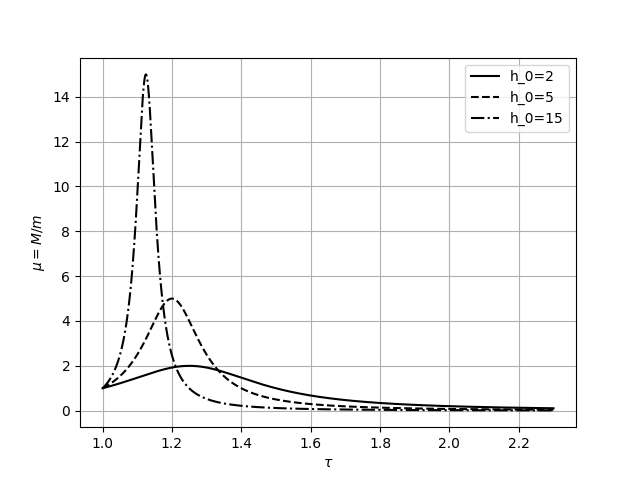}
\caption{Illustration of the behavior of the ratio $\mu \equiv \frac{<M>}{m}$ for the model with $\xi{=}0$ (see (\ref{M33})) as the function of time and of the guiding parameter $h_0$. For illustration on all the figures we put $m{=}1$, $t_*=1$ and measure the cosmological time in the dimensionless units $\tau {=} tH_0$ ($H_0 {=}\sqrt{\frac{\Lambda}{3}}$). The maximal values of all functions $\mu(t,h_0)$ are equal to $h_0$. For $h_0>2m$ the values of $\mu$ grow with $h_0$ and the maxima of the corresponding graphs shift to the initial moment $t_*$.}
\end{figure}

The spinor particle number density is presented as
\begin{equation}
s(t) \equiv \frac{S(t)}{S(t_*)} = \left(\frac{m}{h_0} \right) \frac{\left[1{+} 4h_0^2 \left(t{-}t_{+} \right)^2 \right]}{\cosh^2{\left[\sqrt{\frac{3\Lambda}{4\Gamma}}(t{-}t_*) \right]}}\times
\label{s1}
\end{equation}
$$
\times \frac{1}{\left\{1 {+}\sqrt{1{+}\frac{\kappa}{\Lambda}(\rho(t_0)+ m S(t_0))} \tanh{\left[\sqrt{\frac{3\Lambda}{4\Gamma}}(t{-}t_*) \right]} \right\}^{2}} \,.
$$
This function depends on the parameters $\Lambda$, $\Gamma$, $m$, $h_0$, and on the initial values $\rho(t_0)$, $S(t_0)$, so that the behavior of this function is much more sophisticated than the behavior of $<M>$. However, the numerical analysis shows that the corresponding graphs possess maxima. If we fix $\Lambda$, $\Gamma$, $m$, $\rho(t_0)$, $S(t_0)$ and consider variation of the guiding parameter $h_0$ only, we can state the following: first, the height of the maximum increases, when the parameter $h_0$ grows; second, the maximum of the corresponding graph starts to shift to the late time moments, stops and then moves towards lower values of time. Fig.2 illustrates the details of such behavior.

\begin{figure}
\includegraphics[width=85mm,height=70mm]{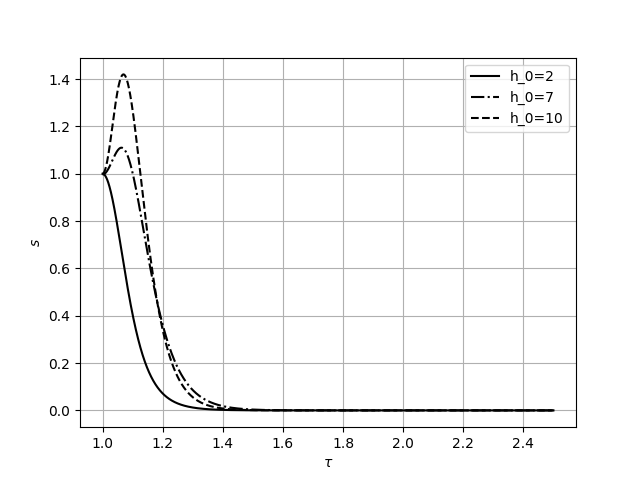}
\caption{Illustration of the behavior of the dimensionless function $s(t)\equiv \frac{S(t)}{S(t_*)}$ for the model with $\xi=0$ (see (\ref{s1})), when the parameters $\Lambda$, $\Gamma$, $m$, $\rho(t_0)$, $S(t_0)$ are fixed, and only the guiding parameter $h_0$ is varying.}
\end{figure}

\subsubsection{The submodel $\xi =\frac12$ and $K=0$}

The condition $\xi =\frac12$ is equivalent to $m_*-m = \frac12 h_0$. The key equation (\ref{XY9}) for the function $X$ reads now
\begin{equation}
\frac{dX}{\sqrt{\frac14 X^2 - \left(\frac{mS(1)}{h_0}\right)^2} } = \mp 4h_0 dt \,.
\label{2XY81}
\end{equation}
The solution, which corresponds to the upper sign in (\ref{2XY81}) and to the condition $h_0>2m$,   is
\begin{equation}
\frac{X(t)}{X(t_*)} =  \cosh{2h_0(t{-}t_*)} {-}\sqrt{1{-}\frac{4m^2}{h_0^2}}\sinh{2h_0(t{-}t_*)} \,,
\label{3XY81}
\end{equation}
or equivalently
\begin{equation}
s(t) = \frac{S(t)}{S(t_*)} = \frac{\cosh{[2h_0(t-t_*)]}}{\cosh^2{\left[\sqrt{\frac{3\Lambda}{4\Gamma}}(t{-}t_*) \right]}}\times
\label{s2}
\end{equation}
$$
\times \frac{\left\{1 {-}\sqrt{1{-}\frac{4m^2}{h_0^2}} \tanh{\left[2h_0(t{-}t_*) \right]} \right\}}{\left\{1 {+}\sqrt{1{+}\frac{\kappa(\rho(t_0)+ m S(t_0))}{\Lambda}} \tanh{\left[\sqrt{\frac{3\Lambda}{4\Gamma}}(t{-}t_*) \right]} \right\}^{2}} \,.
$$
The effective mass evolves with respect to the law
\begin{equation}
<M> = \frac{m}{\cosh{2h_0(t{-}t_*)} {-}\sqrt{1{-}\frac{4m^2}{h_0^2}}\sinh{2h_0(t{-}t_*)}} \,.
\label{XY91}
\end{equation}
At $t=t_*$ we obtain $<M>(t_*) = m$. The function $<M>(t)$ reaches  the maximal value $<M>_{(\rm max)} {=} \frac{h_0}{2} $, where $t_{\rm max}$ is defined as
\begin{equation}
\cosh{[2h_0(t_{\rm max}-t_{*})]}= \frac{h_0}{2m}  \,,
\label{0max3}
\end{equation}
or equivalently
\begin{equation}
t_{\rm max}= t_* + \frac{1}{2h_0}\log{\left[\frac{h_0}{2m}\left(1+ \sqrt{1- \frac{4m^2}{h^2_0}} \right) \right]}   \,.
\label{max3}
\end{equation}
The quantity $t_{\rm max}$ as the function of the guiding parameter $h_0$ increases, when the parameter $h_0$ grows, reaches the maximum and then monotonically tends to $t_*$.  In other words, when the guiding parameter $h_0$ monotonically grows, the maximum of the function $<M>(t)$ drifts to the late time moments, then stops and starts to drift to the initial point $t_*$, as in the submodel with $\xi=0$. Fig.3 illustrates the behavior of the reduced function $\frac{<M>}{m}$ for the cases, when this maximum is passed.

\begin{figure}
\includegraphics[width=85mm,height=70mm]{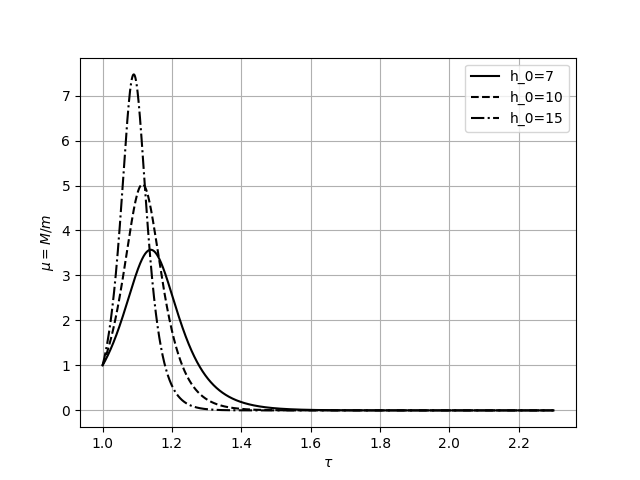}
\caption{Illustration of the behavior of the ratio $\mu \equiv \frac{<M>}{m}$ for the model with $\xi=\frac12$ (see (\ref{XY91})) as the function of time and of the guiding parameter $h_0$. The maximal values of all functions $\mu(t,h_0)$ are equal to $\frac12 h_0$. For $h_0>2m$ the values of $\mu$ grow with $h_0$ and the maxima of the corresponding graphs shift to the initial moment $t_*$.}
\end{figure}

The behavior of the function $s(t)$ (\ref{s2}) is similar to the one for the model with $\xi=0$. Again, the graphs demonstrate the maxima, which move similarly. Fig.4 illustrates this behavior.

\begin{figure}
\includegraphics[width=85mm,height=70mm]{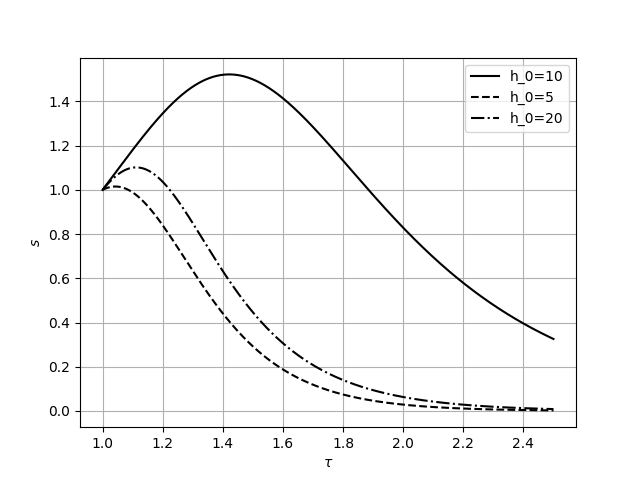}
\caption{Illustration of the behavior of the dimensionless function $s(t)\equiv \frac{S(t)}{S(t_*)}$ for the model with $\xi=\frac12$ (see (\ref{s2})), when the parameters $\Lambda$, $\Gamma$, $m$, $\rho(t_0)$, $S(t_0)$ are fixed, and only the guiding parameter $h_0$ is varying. According to the legend, when $h_0$ grows, the maximum of the graph shifts to the late time moments, then stops and starts to drift towards the initial time moment. The heights of the maxima of the graphs increase on the first stage and decrease on the second one.}
\end{figure}

\subsection{On the solutions for the interval $t>t_{**}$}

During the third stage of the Universe evolution we deal with $\beta=0$, the value of the scalar $S$ on the boundary $t=t_{**}$ is predetermined by the formulas already found for the interval $t_0<t<t_*$. The process of spinorization is finished.

\section{Discussion and conclusions}

{\bf 1. The role of guiding model parameters.}

We presented an exactly integrable phenomenological model according to which the dynamic aether coupled to the spinor field opens a window for the spontaneous growth of the fermion number in the early Universe. 
We have to emphasize that this spontaneous growth is the result of internal self-interaction in the spinor system, which we indicated as a mechanism of self-similar coupling.
As for the dynamic aether, it plays the role of regulator for this process. Our purpose was to show explicitly that the function $S(t)= \bar{\psi} \psi$, which is usually associated with the number
density of the spinor particle, can grow, can reach some maximal value and then monotonically decrease under the influence of the Universe expansion. The formula (\ref{s1}) describes this statement
for the first model with the guiding model parameter $\xi=0$, and the formula  (\ref{s2}) confirms this statement for the model with $\xi=\frac12$. It is required to say a few words about the parameters of the whole model. First, we use the cosmological constant $\Lambda$; second, only one Jacobson coupling constant $C_2$ appears in the master equations reduced for the chosen
spacetime symmetry. Final formulas contain the unified parameter $H_{\infty} = \sqrt{\frac{\Lambda}{3\left(1+ \frac32 C_2 \right)}}$, which gives the asymptotic value of the Hubble function $H_{\infty}$, different from the de Sitter parameter $H_0=\sqrt{\frac{\Lambda}{3}}$.
The model of self-similar interaction includes additional parameters $\xi$ (\ref{XY3}) and $h_0$. It turned out that the parameter $h_0$ predetermines the maximal value of the effective spinor
mass $<M>$ (\ref{D17}). Also we introduced phenomenologically two time moments $t_*$ and $t_{**}$, which restrict the interval inside of which the dynamic aether "allows" the spinor field to switch
on the self-similar interaction; we consider the corresponding values of the expansion scalar $\Theta_* \equiv \Theta(t_*)$ and $\Theta_{**} \equiv \Theta(t_{**})$ as the analogs of the first and
second Curie temperatures in ferroelectrics. The last guiding parameter of the model is the so-called "seed mass" $m$. This value of the spinor mass is not fixed in the model. there are several ideas about this quantity. For instance, the collision of pairs of photons could produce the electron-positron pairs, when the photon energy  was sufficient  to overcome the barrier of the electron
rest energy $2m_e$. In this case the electron mass $m_e$ could be the seed mass $m \to m_e$. The corresponding initial fermion number density was indicated as $S(t_0)$.

{\bf 2. The role of the effective spinor mass.}

In the thermodynamics of the hot Universe there exist a hypothesis that when the thermal energy $k_B T$ is equal to the sum of the rest energies of pair of particles, $k_B T= 2 {\cal M} c^2$,
these particles can emerge as the individual ones (particles drop out from the equilibrium fluid). We put forward a hypothesis that in analogy with the thermodynamic approach, the specific spinor
particles can appear as the individual ones, when their masses (predicted by the quantum theory) coincide with the effective mass $<M>$. Is it possible on the basis of our hypothesis to explain the
birth of spinors of all known types?

In the standard units we have to replace $m \to \frac{mc}{\hbar}$. For the special units with
$c=1$ and $\hbar=1$ this mass parameter is presented as usual in $MeV$. From the standard catalog of fermion masses we know the following: first, the masses of the quarks
are in the range between $174340 \pm 790$ MeV (for the t-quark) and $1.5 \leftarrow \rightarrow 5$ MeV (for the u-quark); second, the masses of protons and neutrons are $938.272$ MeV and $939.565$ MeV, respectively; third, the masses of leptons are in the range between $1776.99$MeV (for the $\tau$-lepton)
and $0.511$ MeV (for the electron); fourth, the masses of the neutrinos are estimated to be in the range between $<15.5$ MeV (for the $\tau$-neutrino) and $<0.0000022$ MeV (for the electron neutrino).
(The masses of the anti-particles coincide with the ones of the corresponding particles).

Based on the solutions obtained for two exactly integrable models we can notice that the effective spinor mass $<M>$ as the function of cosmological time starts from the value $m$ at $t=t_*$,
reaches the maximal value ($h_0$, if $\xi=0$ and $\frac12 h_0$, if  $\xi = \frac12$, and then tends to zero.
In fact, we can estimate the parameter $h_0$ associated with the maximal value of the effective mass, as $h_0 > 174340 \pm 790$ MeV for the model with $\xi=0$, and as
$h_0 > 348681 \pm 580$ for the model with $\xi=\frac12$. The mass of the electron neutrino, the minimal mass from this catalog, points on the moment of time $t_{**}$, when the self-similar interaction was switched off by the dynamic aether.

{\bf 3. On the maximal spinor particle number density.}

 The idea of spontaneous spinorization assumes that during the interval of the cosmological time $t_* <t<t_{**}$ a significant growth of the spinor number density $S$ takes place. This idea
 is confirmed by the exact solutions (\ref{s1}) (for $\xi=0$) and (\ref{s2}) (for $\xi=\frac12$). Illustrations presented on Fig.2 and Fig.4 visualize the growth of the function $S(t)$. It is important to notice that the maximal value $S_{\rm max}$ is predetermined by the model guiding parameter $h_0$. If we suppose that the model with $\xi =0$ is appropriate,
 the estimation gives that $S_{\rm max} \propto S(t_*) 4m h_0 (t_{\rm max}{-}t_{+})^2$; for the hypothetical value $h_0 > 174340 \pm 790$ MeV it is rather big quantity. If the model with
 $\xi = \frac12$ is more appropriate, the estimations give $S_{\rm max} \propto S(t_*) \cosh{2h_0(t_{\rm max} {-} t_*)}$.

{\bf 4. What is the energy source for the spontaneous spinorization?}

We think that the energy required to increase the number of fermions is drawn from the energy reserve of the gravitational field. The presence of the term $m S {+} \beta $
in the right-hand side of the equation (\ref{keyGravity}) hints us that the energy can be effectively redistributed between the gravitational and spinor fields, when the aether opens a window for this process.

{\bf 5. Outlook.}

We hope to apply the presented results to the realistic cosmological model, however, this work is outside the scope of this article; we hope to organize the detailed analysis in the next paper.

 \acknowledgments{The work was supported by the Russian Science Foundation (Grant No 21-12-00130).}

\end{document}